\documentstyle[aps,prl,multicol,graphicx]{revtex}

\begin{document}
\input{epsf}
\draft
\tighten

\preprint{}
\title{InAs-AlSb quantum wells in tilted magnetic fields}
\author{S. Brosig and K. Ensslin}
\address{Solid State Physics Laboratory, ETH Z\"{u}rich, 8093
Z\"{u}rich,  Switzerlandx}
\author{A. G. Jansen}
\address{High Magnetic Field Laboratory, MPIF-CNRS, B. P. 166, 38042
Grenoble Cedex 9, France}
\author{C. Nguyen, B. Brar, M. Thomas, and H. Kroemer}
\address{Department of Electrical and Computer Engineering, University of
California, Santa Barbara, Ca 93106, USA}
\date{\today}
\maketitle
\begin{abstract}  
  InAs-AlSb quantum wells are investigated by transport experiments in
  magnetic fields tilted with respect to the sample normal. Using the
  coincidence method we find for magnetic fields up to 28 T that the
  spin splitting can be as large as 5 times the Landau splitting. We
  find a value of the $g$-factor of $\vert g \vert \approx 13$. For
  small even-integer filling factors the corresponding minima in the
  Shubnikov-de Haas oscillations cannot be tuned into maxima for
  arbitrary tilt angles. This indicates the anti-crossing of
  neighboring Landau and spin levels.  Furthermore we find for
  particular tilt angles a crossover from even-integer dominated
  Shubnikov-de Haas minima to odd-integer minima as a function of
  magnetic field.
\end{abstract}


\begin{multicols} {2}
\narrowtext
\def\rxx{$\rho_{xx}$}
\def\deg{$^\circ$}
\section{Introduction}
The energy spectrum of two-dimensional electron gases (2DEG) in
magnetic fields of arbitrary orientation is fairly well understood
\cite{Ando1982,Maan1984}. Most considerations follow a single-particle approach
which is powerful to explain several of the experimentally observed
features. For magnetic fields tilted with respect to the sample normal
one finds that the Landau splitting, which is proportional to the
component of the field perpendicular to the 2DEG can be tuned with
respect to the Zeeman splitting which is proportional to the total
magnetic field.  This is used in the so-called coincidence method
\cite{Fang1968} where appearance and disappearance of minima in
Shubnikov-de Haas oscillations (SdH) as a function of tilt angle is
observed in magnetotransport experiments. The analysis in terms of a
picture of non-interacting electrons has proven very powerful for the
analysis of energy spectra in Si-MOSFETs \cite{Fang1968}, InAs-GaSb
superlattices \cite{Chang1982}, InAs-GaSb quantum wells
\cite{Smith1987}, GaAs-AlGaAs heterostructures \cite{Nicholas1988},
GaInAs/InP heterostructures \cite{Koch1993} and Si/SiGe
heterostructures \cite{Weitz1996}.  In this paper we focus on
InAs-AlSb quantum wells and extend preliminary studies on this
material system \cite{Brosig1998}. We present several features that
are perfectly well explained in the existing single-particle picture,
namely 1. the appearance and disappearance of even- and odd-integer SdH
minima as a function of tilt angle, 2. a Zeman splitting as large as
five times the Landau splitting for tilt angles around 87\deg, and 3.
a g-factor for InAs of about 13 in agreement with considerations based
on conduction band non-parabolicity \cite{Roth1959}.  In contrast to
straight forward expectations we find 4. non-vanishing SdH minima for
even-integer filling factors $\nu = 4,6,8$ in the range of tilt angles
and magnetic fields where these filling factors can be observed and 5.
a regime at low magnetic fields where even-integer filling factor SdH
minima persist for all tilt angles, while the usual coincidence
features occur at higher magnetic fields.  These observations are
discussed in view of other experiments \cite{Koch1993,Weitz1996} and
theoretical ideas based on exchange enhancement
\cite{Giuliani1985,Yarlagadda1991}.

\section{Level crossing in the single particle regime}

All samples contained 15nm wide InAs quantum wells, confined by AlSb
or Al$_{x}$Ga$_{1-x}$Sb $(x \le 0.8)$ barriers.  The sample details
are summarized in \cite{Tuttle90,Brosig1999}.  Our samples are of very
high quality and have mobilities up to 84\,m$^2$/Vs. In this paper we
focus on a sample with a GaSb cap and a carrier density of
$N_s = 6.2 \cdot 10^{11}\,$cm$^{-2}$ (UCSB
\#~9503-18).  
The samples were patterned into geometries suitable for transport
experiments and equipped with Ohmic contacts to the 2DEG.

The samples were mounted on a revolving stage in several cryostat
environments.  The angle $\alpha$ is measured between the magnetic
field orientation and the sample normal.  For the data taken at
T=1.7\,K and magnetic fields up to 8\,T the revolving stage was
computer controlled. Consequently very dense data sets were obtained.
We also measured the samples in a dilution refrigerator at sample
temperatures down to 100 mK and magnetic fields up to 15 T as well as
in a $^3$He system with a base temperature of about 400\,mK and
magnetic fields up to 28\,T.  Because of the Landau level broadening
the temperature dependence of the SdH oscillations basically levels
off below 1.7\,K. The difference in experimental resolution of the
three setups is mostly determined by the respective measurement
electronics.

The results obtained on different samples depend on the carrier
concentration. The filling factor is
definded by $\nu=N_sh/eB$, where $N_s$ is the electron density of the
2DEG.  For perpendicular fields, i.e. $\alpha=0$, all observed
features at magnetic fields $B \le 1.5$\,T, where the Zeeman splitting
is not yet resolved, can be analyzed with one single SdH period with
very high accuracy \cite{Brosig1999}. Effects of inversion asymmetry
induced zero-field spin-splitting \cite{Rashba60} are therefore not
considered.  From the largest filling factors that  we can observe we
estimate the Landau level width to about 0.4\,meV.

We can follow the disappearance and reappearance of minima at even- and 
odd-integer filling
factors as a function of increasing
tilt angle. This interplay between
pronounced even- and odd-integer filling factor minima occurs for a
series of angles. It also shows up in the respective quantum Hall
plateaus \cite{Brosig1998}.

Figure 1 shows magnetoresistance traces at specific tilt angles where either
SdH minima occur only at even-integer filling factors, at even and odd,
or only at odd integer filling factors. The amplitude of the SdH oscillations
at large tilt angles is magnified with respect to the other traces.
The angles are determined by measuring $\rho_{xy}$ with high accuracy.
The absolute error in the angle becomes larger with increasing tilt angle
because of the $\cos(\alpha)$-dependence.

We find that the carrier density decreases by up to 5\%
if parallel magnetic fields larger than 20\,T are applied. This also
shows up in a non-linear Hall effect for large parallel fields. We
attribute this behavior to magnetic freeze-out of carriers due to a
redistribution of the electrons from the well into some localized
states. The reason for this could be a strong diamagnetic shift of the
quantum well state. This effect has no consequences for the results
presented in this paper but explains why the SdH minima in Fig.~2 for
large tilt angles $\alpha$ do not exactly fall onto the dashed lines.

The inset in Fig.~2 describes the various coincidence situations which
are characterized by the parameter $r$, the ratio of Zeeman and
cyclotron energies.
$$r=\frac{g \mu_{\rm B}B_{tot}}{\hbar \omega_c}$$
Here
$\omega_c=eB_{\perp}/m^*$, $m^*$ is the effective electron mass,
$\mu_{\rm B}$ is the Bohr magneton and $B_{\perp} = \cos(\alpha) \cdot
B_{tot}$. We thus arrive at
$$r \cdot \cos(\alpha) =\frac{gm^*}{2m_e},$$
where $m_e$ is the free
electron mass.  The data in Fig.~1 shows the resistance traces at
$r$-values always close to the indicated numbers of 1/2, 1,
3/2,\ldots.  The larger the tilt angle, the more difficult it is to
realize a given coincidence situation accurately since the span of
angles at which it takes place decreases with $\cos\alpha$.
Nevertheless we demonstrate that SdH oscillations can be measured in a
situation where the Zeeman splitting is 5 times larger than the Landau
splitting.

Figure~2 shows the coincidence situations plotted as $1/\cos(\alpha)$
versus $r$. The slope of this curve is proportional to the product
$gm^*$. We determined the effective mass for this sample by
temperature dependent SdH measurements and found a value for the
effective mass of $m^*=(0.032 \pm 0.002) \cdot m_e$ which is in
agreement with values reported in the literature. Using this value for
$m^*$ we computed $\vert g \vert \approx 13$.

Such experiments have been performed on a series of samples. In first
approximation the obtained data can be described by using Landau
levels and spin levels behaving and crossing as expected in a
single-particle model. Because the $g$-factor is so large effects of
electron-electron interactions, the so-called exchange enhancement
\cite{Ando1974}, are expected to be relatively small. Furthermore
these effects should increase for decreasing filling factors. In our
case the experimental data can best be described with a product $gm^*$
which is constant over the investigated range of magnetic fields and
angles.

The effects of non-parabolicity can be estimated using a $\bf{k} \cdot
\bf{p}$ formalism \cite{Kane1957} which in its simplest case reduces
to the two-band model.

$$m^*(E)=m^*(E=0) \cdot \left(1+2 \frac{E}{E_{\rm g}}\right)$$

Here $E_{\rm g}=400$\,meV is the band gap of InAs and $E$ is the
electron energy relative to the conduction band edge. Because
of the huge conduction band offset between InAs and AlSb (1.35\,eV), we
use the model of a quantum well with infintely high walls. The total
energy $E$ can, to a good approximation, be written as the sum of an
approximate Fermi energy 
$E_{\rm F}=N_{s} \cdot \pi\hbar^{2} / m^{*}$ 
and an approximate confinement energy 
$E_{c}= \hbar^{2}/2m^{*}\cdot \pi^{2}/a^{2}$,
where $a$ is the quantum well width. With this we obtain for the
density dependence of the effective mass in the two-band model  
$$ m^{*}(N_{s}) = \frac{m^{*}_{0}}{2} + \frac{m^{*}_{0}}{2} 
\sqrt{1+\frac{8}{E_{\rm g}} \left( \frac{\hbar^2}{2m^{*}_{0}}
\frac{\pi^2}{a^2}
+ \frac{\pi\hbar^2}{m^{*}_{0}} N_{s} \right) } $$
Here $m^{*}_{0} = m^{*}(E=0)$, i.e. the effective mass at the 
conduction band edge, which for InAs is $m^*_{0}/m_e=0.023$.
We find $m^*(N_{s}=4.4 \cdot 10^{11}\,$cm$^{-2})/m_e=0.032$ in agreement with
our experimentally determined value. The values for the energies are 
$E_{\rm F}= 52$\,meV and $E_{c}=51.6$\,meV. 

At the same time the $g$-factor is reduced \cite{Roth1959}
in agreement with our experimental findings.
For the $g$-factor the two-band model results in

$$g(E)=g(E=0) \cdot (1- \alpha \cdot E)$$

The parameter $\alpha$ is estimated in Ref. \cite{Scriba1993} to
be $\alpha = 0.0025 \cdot $1/meV for a quantum well system very similar to
ours. This results in a $g$-factor of $\vert g \vert =12$ very close to our
experimental result. Using the expressions for the $g$-factor and the
effective mass one finds that the total effect of nonparabolicity on
the product $gm^*$ almost cancels out.

Several additional aspects should be considered in this discussion.
For large tilt angles the in-plane magnetic field component can be as
large as 10\,T.  In this case it is well known that the Fermi surface
is no longer a circle but an ellipse. The effective mass thus depends
on $B$ \cite{Maan1984}.  We measured the temperature dependence of the
SdH oscillations in tilted magnetic fields in order to extract the
effective mass as a function of field and tilt angle. Within the
experimental accuracy we found that the effective mass is constant to
5\% in the investigated parameter regime. On the same footing one also
expects that the $g$-factor becomes a magnetic field dependent quantity.
With these complications in mind one has to take the analysis of the
product $gm^*$ from the plot in Fig.~2 with a grain of salt.

\section{Level anti-crossings at small filling factors}
Figure~3 shows magnetoresistance traces down to even-integer filling
factors of $\nu=6$. We only present the range of angles where the
situation corresponding to $r=1$ occurs. The tilt angle is changed in
rather small increments which are monitored by the change in the Hall
resistance $\rho_{xy}$. Similar but less pronounced features
alos occur in a sample has a lower
carrier density of $N_s = 4.4 \cdot 10^{11}\,$cm$^{-2}$. The
highest perpendicular magnetic fields correspond to total magnetic
fields of 28\,T. For $\alpha = 73.5$\deg, minima occur for even- and
odd-integer filling factors. As the tilt angle increases, even-integer
minima weaken until about $78.8^\circ$ and then increase again in
strength. They never completely disappear even up to filling factors
of $\nu$=16. This means that there always remains a minimum of the density
of states at the Fermi energy when the single-particle model predicts
a crossing of spin and Landau levels.

An anti-crossing of single particle levels has been predicted for
filling factor $\nu=2$ \cite{Giuliani1985,Yarlagadda1991} based on the
transition from a spin-unpolarized state at small tilt angles to a
spin polarized state at large tilt angles. Experimental data obtained
on GaInAs/InP heterostructures \cite{Koch1993} showed the expected
single particle behavior for low-mobility samples while a
non-suppression of the SdH minimum at $\nu=2$ for high-mobility
samples was observed. This was interpreted in the framework of the
formation of a spin-polarized ground state \cite{Giuliani1985} induced
by the strong parallel magnetic field. In the case of Ref.
\cite{Koch1993} the SdH minimum corresponding to filling factor
$\nu=4$ and higher even-integer filling factors were perfectly well
suppressed at the same tilt angle as expected in a single particle
model. The authors \cite{Koch1993} argued that for low mobility
samples and higher integer filling factors neighboring levels overlap
due to their broadening and the exchange interaction cannot help to
further open the gap.

The experimental situation in our case is different in the following
aspects. The SdH minima at even-integer filling factors weaken but
do not disappear. Furthermore their weaking goes hand in hand with
their overall appearance, i.e.\ the sudden importance of an exchange
driven opening of a gap cannot be observed. Unfortunately the carrier
density in our samples is too high to observe the behavior of SdH
minima corresponding to filling factors $\nu=2$ and $\nu=4$ at large
tilt angles and experimentally accessible magnetic fields.

In order to get an understanding of the energy struture in tilted
magnetic fields we calculated the magnetoresistance following
Gerhardts \cite{Gerhardts1976}. We included a constant background
density of states in order to model the broad minima in the
magnetoresistance.  Based on the single particle energies
$$E_{s,n}=\hbar\omega_c\left(n+\frac12\right)+s\cdot g\mu_{\rm B} B,
n=0,1,2,\ldots, s=\pm\frac12$$
an anti-crossing between
neighboring levels of $\Delta E=0.29 \hbar\omega_c$ was inserted in
the model. At $B_{\perp}$=4.2\,T ($\nu=6$) and $m^{*}=0.032 \cdot m_{e}$, this
corresponds to $\Delta E=4.4$\,meV.  We assumed a Gaussian Landau
level broadening $\Gamma = \hbar / \tau_{q}=1.5\,$meV with
$\tau_{q}=0.45$\,ps.

The magnetic field dependence of the anti-crossing was approximated
with a smooth parabolic curvature.

Figure~4 shows calculated resistance traces. There is at least
qualitative agreement between the calculated (Fig.~4) and experimental
(Fig.~3) data sets. From the simulation it is obvious that the
situation where even-integer minima in the SdH oscillations are
weakened or even suppressed extends over a significantly larger range
of angles compared to the experiment. This could arise from our rough
modelling but also hints at the importance of interaction effects for
the details of SdH oscillations.

What could be the reason for the persistent appearance of even-integer
SdH minima in the regime where the underlying single particle energy
levels are expected to cross? For small filling factors the effects of
exchange enhancement \cite{Ando1974} have been demonstrated in various
experiments (for a recent example see \cite{Leadley1998} and
references therein). From our experimental data at small filling
factors we do not see an indication that eletron-electron interactions
in terms of exchange enhancement play a significant role. For the case of 
perpendicular magnetic fields, $\alpha=0$, the energy levels are described by
three quantum numbers, namely subband, Landau and spin quantum
numbers. This is based on the fact that the Hamiltonian can be
separated into a part describing the electron motion in the plane of
the 2DEG and another part responsible for the quantization in growth
direction. For tilted magnetic fields mixed levels arise whose degenracy 
is still completely controlled by the perpendicular magnetic field 
component \cite{Maan1984}. For the InAs-AlSb system this approach has to be extended
in order to incorporate the strong conduction band non-parabolicity of
InAs, as well as the possible strain in the well due to the different lattice
constants of barrier, well and GaAs substrate. One can envision that
such effects already lead to possible level couplings and
anti-crossings as observed in the experiment.

\section{Even-integer SdH minima at low magnetic fields}
Figure~5 presents a grey-scale plot composed of magnetoresistance
traces taken at very closly spaced tilt angles around the regime of
$r=1$ and $r=2$. Here we focus on the regime of small magnetic fields. For
$\alpha=65^\circ$ SdH minima occur at even-integer filling factors. As
the tilt angle is increased, odd-integer minima take over at magnetic
fields $B_{\perp} \ge 0.8$\,T and gradually disappear again in favor of
even-integer minima. At magnetic fields below 0.8\,T minima occur only
at even-integer filling factors over the whole range of tilt angles.
The inset of Fig.~5 shows a representative resistance trace at an
intermediate tilt angle where the SdH oscillations are dominated by
even-integer minima at low magnetic fields, a crossover regime and odd
integer filling factors at higher magnetic fields. A beating pattern
would not display such a phase shift in the pattern of the
oscillations.

Starting from the Landau and spin levels in tilted magnetic fields
such a behavior can occur in two ways: either the Landau energy is not
exactly proportional to the perpendicular component of the magnetic
field, or the Zeeman splitting is angle dependent. Both effects have
been discussed to some extent before. Non-parabolicity effects are
most likely a minor contribution for such small magnetic fields. The
large in-plane magnetic field component, which can lead to an
anisotropic effective mass dispersion, should become more important
for larger magnetic fields. However, the unusual behavior as presented
in Fig. 5 occurs in the low-magnetic field regime.

Leadley et al. have shown that there is a critical collapse of the
exchange enhanced spin splitting in two-dimensional systems
\cite{Leadley1998}. The authors found that the total spin splitting is
a sum of the bare Zeeman splitting proportional to the total magnetic
field and a contribution due to exchange enhancement which is
proportional to the perpendicular component of the magnetic field.

$$\Delta_{spin}=g_{0}\mu_{\rm B}B_{tot}+\beta \hbar e B_{\perp}/m^*$$

For the case of GaAs heterostructures, Leadley at al. found $\beta
=0.2$ independent of magnetic field. In their case $g_{0}$ is the bare
$g$-factor because non-parabolicity effects are negligible in GaAs. In
our case $g_{0}$ has to be identified with $g(E)$ where the
non-parabolicity contribution stems from the position of the Fermi
energy above the conduction band edge and does not depend on magnetic
field in the investigated range of parameters.

In the regime of large magnetic fields discussed before, where spin 
splitting is well resolved, we found that
the exchange enhancement is a minor contribution. However, for small magnetic
fields and large tilt angles the exchange contribution could play an
important role. If the bare spin splitting is smaller than the Landau
level broadening, the exchange enhancement is not expected to play a
role. In this case even-integer SdH minima will dominate the
magnetoresistance for all tilt angles. Once the bare Zeeman splitting
approaches and exceeds the Landau level broadening the exchange enhancement will
further increase the spin gap and the usual coincidences between
Landau and spin levels will take over.

For any functional dependence of g on B which is smooth one would not 
expect a sudden crossover from even-integer to odd-integer minima as 
depcited in the inset of Fig. 5. The sudden change in periodicity over a 
small magnetic field range requires a mechanism which leads to an abrupt 
opening of the spin gap similar as it has been observed in 
Ref. \cite{Leadley1998} for the cititcal collapse of the exchange enhanced 
spin-splitting.

\section{Summary}
We have presented a series of SdH measurements on InAs-AlSb quantum
wells in tilted magnetic fields. In a reasonable range of parameters
the experimental results can be understood in a straight forward
single particle model. The coincidence method is based on independent
Landau and spin levels.  This way we obtain reasonable numbers for the
effective mass and g-factor that agree with results of a two-band
model and experimental results of others. For large magnetic fields we
find an anti-crossing of neighboring Landau and spin levels. Most
likely this is not a consequence of electron-electron interactions. We 
speculate that this effect
arises from the pronounced non-parabolicity of the InAs conduction
band as well as from the built-in strain in such samples. For very
small magnetic fields SdH minima exist only at even-integer filling
factors independent of tilt angle. This is attributed to a critical
filling factor necessary for the observation of spin-splitting

We are grateful to R. Warburton and S. Ulloa for helpful discussions and
thank ETH Z\"urich and QUEST for financial support. The hospitality 
of the High-Magnetic Field Laboratory in Grenoble is gratefully 
acknowledged.

\begin{figure}
    \caption{\rxx-traces for various tilt angles where different
      coincidence situations are met. The topmost three curves are scaled
      up $\times$5 and a parabolic background has been subtracted for
      clarity.}
    \label{Fig1}
\end{figure}

\begin{figure}
    \caption{Coincidence plot gathered from the angles in
      Fig.~1. The straight line has a slope of 4.8, yielding
      $g=2/({\mbox\rm slope}\cdot0.032)=13$, where the effective mass
      $m^*/m_e=0.032$ has been unsed.}
    \label{Fig2}
\end{figure}

\begin{figure}
    \caption{Magnetoresistance $\rho_{xx}$ versus the perpendicular 
      component of magnetic fields for various tilt angles in the
      regime of coincidence r=1, $g\mu_{\rm B}B=\hbar\omega_{\rm c}$.
      Non-vanishing minima at even-integer filling factors are
      observed.}
    \label{Fig3}
\end{figure}

\begin{figure}
    \caption{Model calculation of the magnetoresistance for various
      tilt angles. The parameters have been chosen to match the
      experimental data presented in Fig. 4. The bold curve is at the
      angle where the coincidence $\hbar\omega_c=g\mu_{\rm B}B$ is met}
    \label{Fig4}
\end{figure}

\begin{figure}
     \caption{Grayscale plot of \rxx\ data. A slowly-varying background has
      been removed and the oscillation amplitude has been raised at low
      magnetic fileds to make the effect visible. The vertical axis
      is linear in $1/\cos(\alpha)$. Black (white) areas indicate
      small (large) values of the resistance. The insert shows one
      (unprocessed) \rxx\ curve at $\alpha$ = 78.2$^\circ$ (Horiz. line in
      grayscale plot). The triangles mark the positions of even-integer 
      filling factors.}
    \label{Fig5}
\end{figure}


\end{multicols}

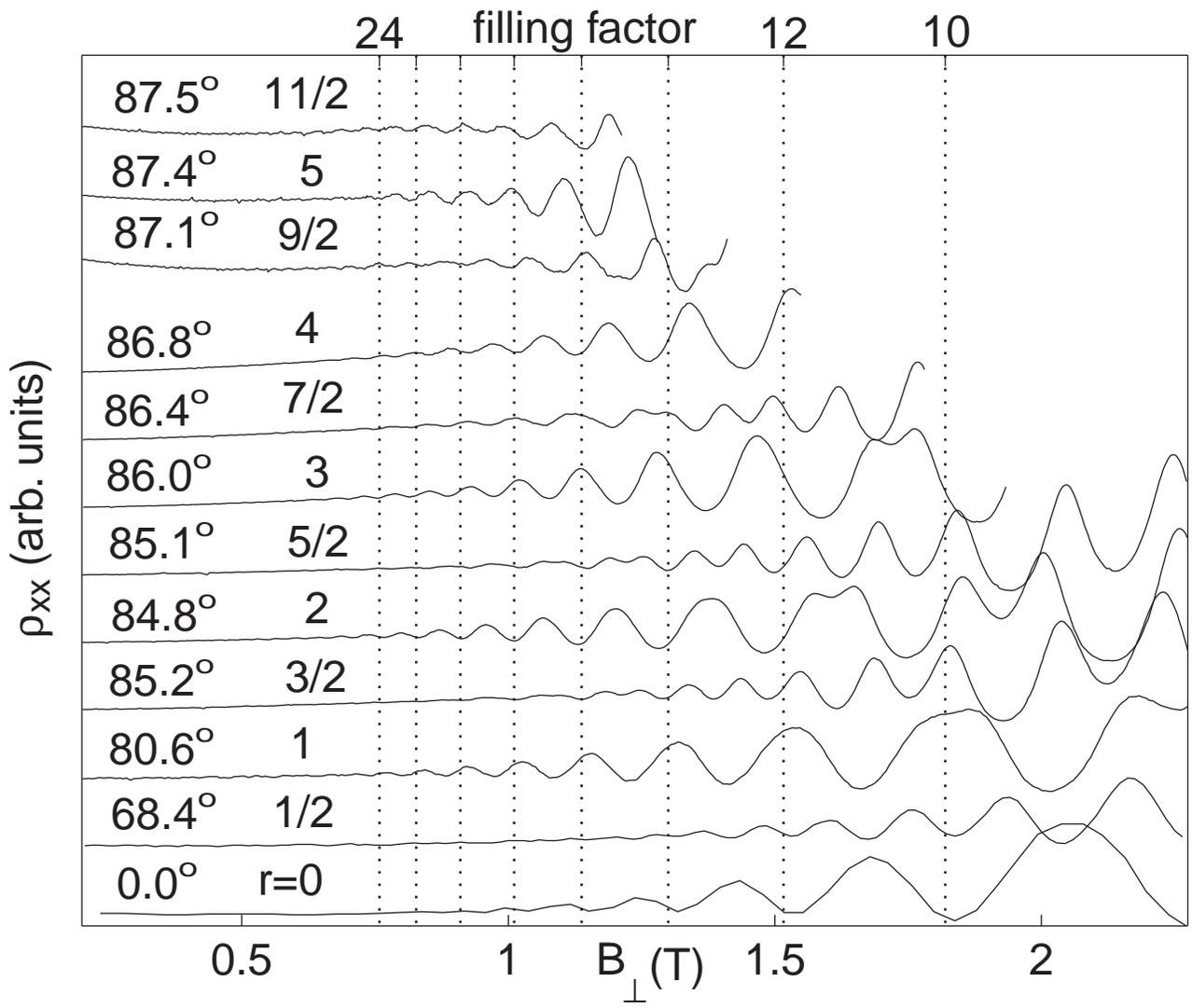
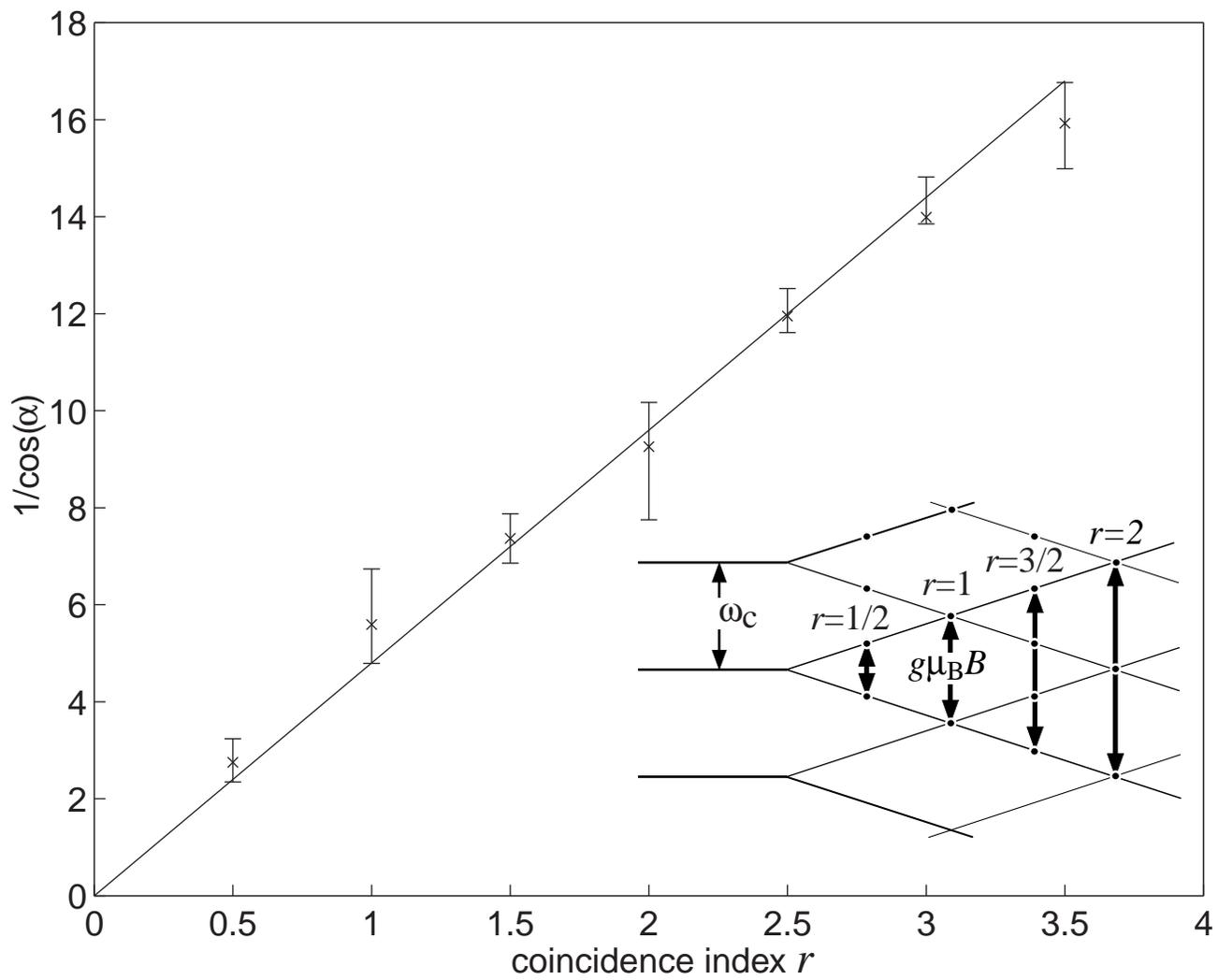
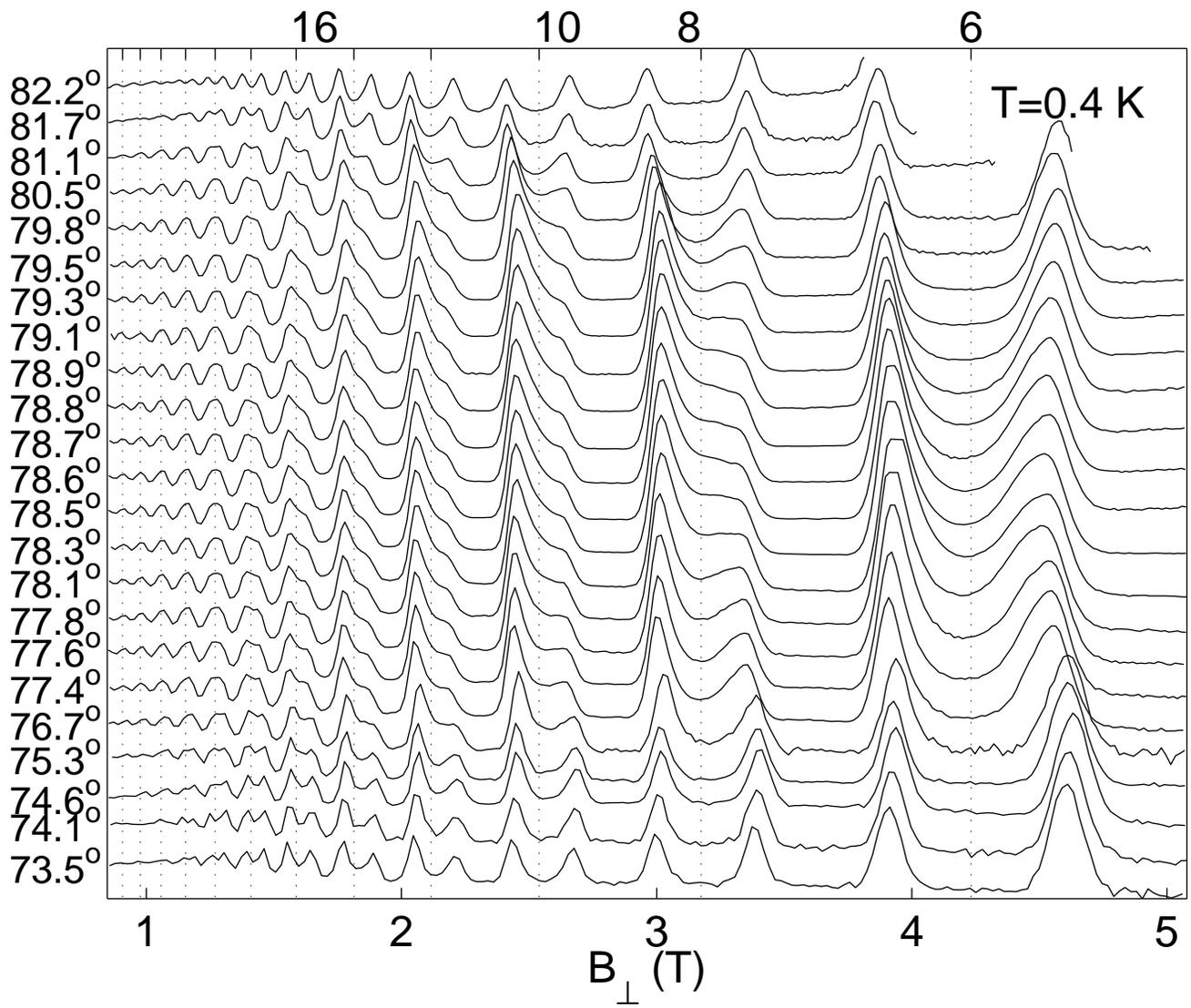
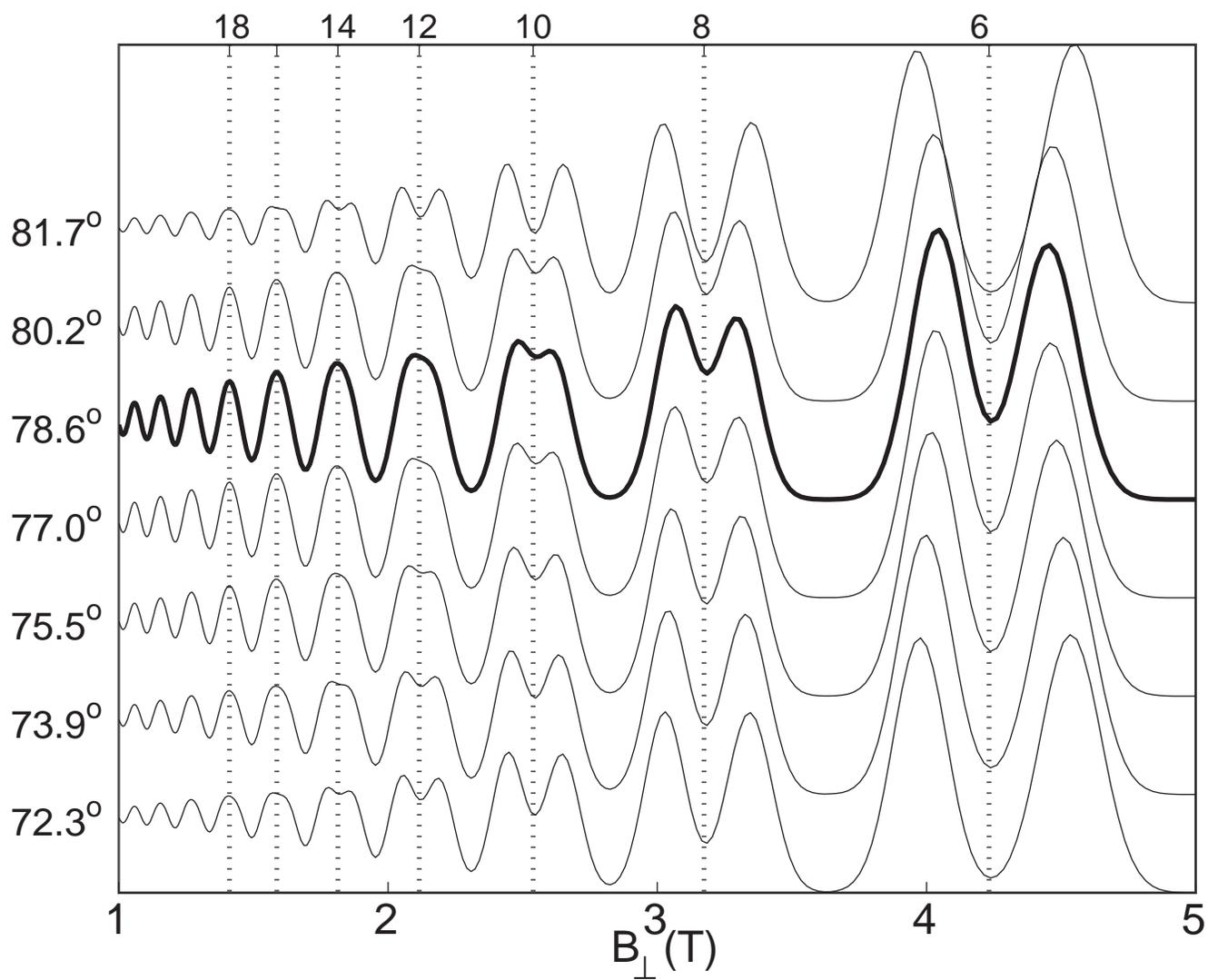
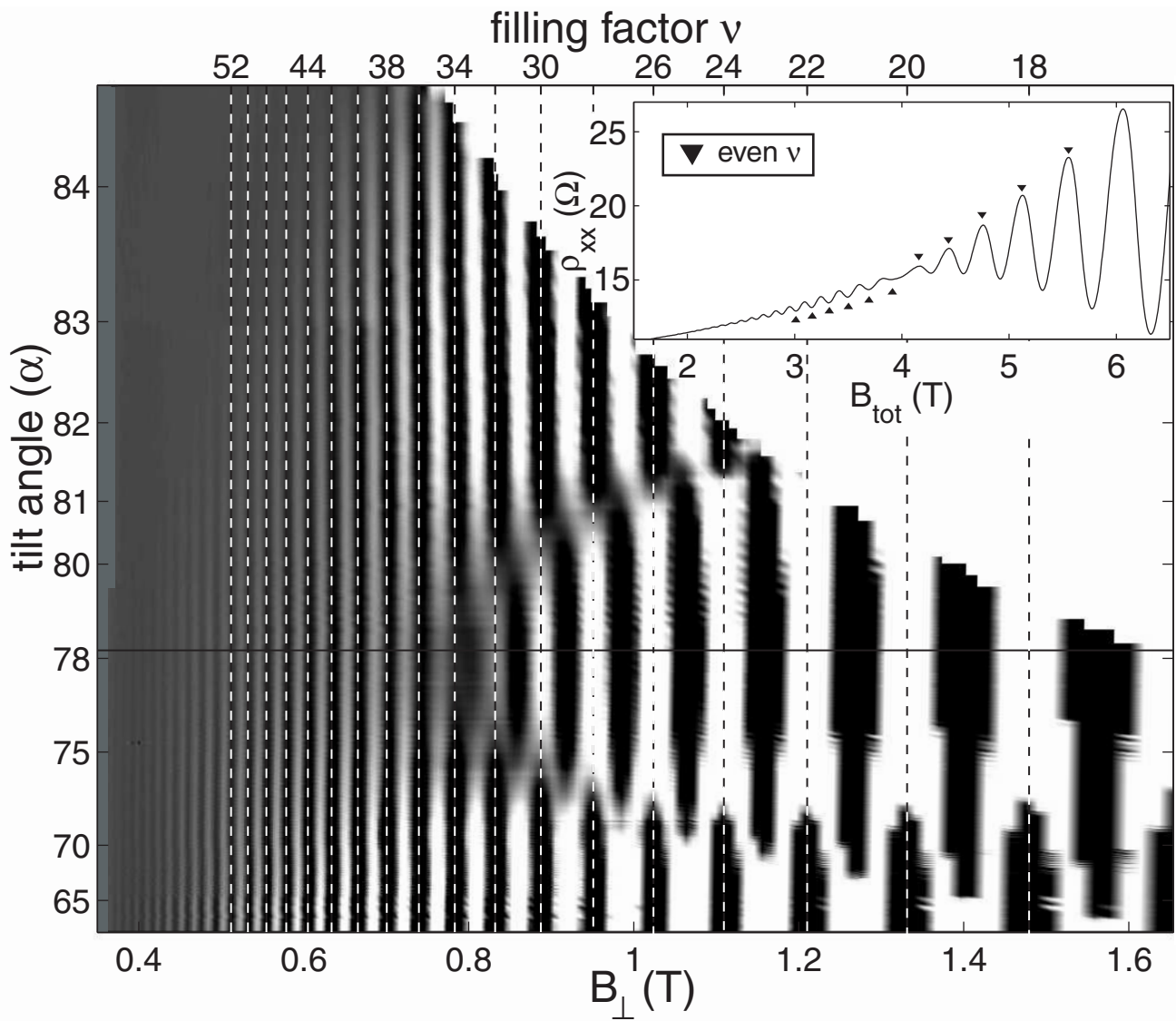
\end{document}